# The Effect of the Multi-Layer Text Summarization Model on the Efficiency and Relevancy of the Vector Space-based Information Retrieval

Ahmad Hussein Ababneh
University of Huddersfield, UK
Ahmad.ababneh@hud.ac.uk

Joan Lu
University of Huddersfield, UK
j.lu@hud.ac.uk

Qiang Xu
University of Huddersfield, UK
q.xu2@hud.ac.uk

*Abstract*-**The massive upload of text on the internet creates a huge inverted index in information retrieval systems, which hurts their efficiency. The purpose of this research is to measure the effect of the Multi-Layer Similarity model of the automatic text summarization on building an informative and condensed invert index in the IR systems. To achieve this purpose, we summarized a considerable number of documents using the Multi-Layer Similarity model, and we built the inverted index from the automatic summaries that were generated from this model. A series of experiments were held to test the performance in terms of efficiency and relevancy. The experiments include comparisons with three existing text summarization models; the Jaccard Coefficient Model, the Vector Space Model, and the Latent Semantic Analysis model. The experiments examined three groups of queries with manual and automatic relevancy assessment. The positive effect of the Multi-Layer Similarity in the efficiency of the IR system was clear without noticeable loss in the relevancy results. However, the evaluation showed that the traditional statistical models without semantic investigation failed to improve the information retrieval efficiency. Comparing with the previous publications that addressed the use of summaries as a source of the index, the relevancy assessment of our work was higher, and the Multi-Layer Similarity retrieval constructed an inverted index that was 58% smaller than the main corpus inverted index.**

*Index Terms_*Automatic Text Summarization, Inverted Index, Jaccard Coefficient, Latent Semantic Analysis, Multi-Layer Similarity, Vector Space Model.

## Abbreviation

ATS:     Automatic Text Summarization
CR:      Condensation Rate
IR:      Information Retrieval
JCM:     Jaccard Coefficient Model
LSA:     Latent Semantic Analysis
MLS:     Multi-Layer Similarity
VSM:     Vector Space Model

## I. INTRODUCTION

Information Retrieval (IR) is the field of computer science that establishes a framework on how to store, represent, and retrieve text documents from a massive collection of unstructured text (i.e., Internet). The IR process includes a series of operations that include text processing, index creation, statistical weighting and matching, text retrieval, and documents ranking. IR plays an essential role in all aspects of life, and any improvements achieved on the IR makes the extracted information from the internet more relevant.

One of the major problems facing the researchers in the IR field is the massive growth of the text volume on the internet. In 2018, the worldwide web-size site [1] published that the number of indexed pages on the internet reached 5.28 billion pages. This huge volume of text imposes to find an innovative solution to store and retrieve the text data efficiently.

IR aims to retrieve relevant documents that match the user information need [2]. The IR system accepts two kinds of data, the user query that reflects the user information need, and a collection of a huge number of documents stored in the inverted index. The inverted index is a data structure that represents the contents of the document as an ordered list of weighted stems or roots (Tokens) [2]. On the user search, the inverted index is inspected to make the necessary matching between the user terms and the index terms. Therefore, a well-established index is necessary to optimize the retrieval speed and performance [3].

Two problems are associated with the use of the inverted index in the IR systems: the space it occupied on the disk or in the computer memory, and the insertion, deletion, and update operation [4], [2]. In this paper, we try to remedy the space overhead of the inverted index. Lin and Chris in [5] stated that the space overhead of the inverted index varies and unpredictable. To solve the problem of the space overhead, we proposed to reduce the size of the original documents before the indexing step is initiated, and this reduction can be performed by extracting the salient components of the documents using efficient and accurate text summarization system.

Automatic Text Summarization (ATS) is a computerized process of condensation which yields a shorter version of the original text and keeps the core meaning and the main ideas reserved [6]. To solve the problem of text data overload, we enhanced the solutions proposed in [7], [8], and [9]. The authors of those solutions tried to reduce the inverted index by using the document summaries that are generated automatically by the ATS systems. The summaries are used as the source of the index instead of the original documents. Two reasons impede the use of ATS as a supporting tool to enhance information retrieval performance.

- The summarization method: in [7], [8], and [9], the authors used statistical techniques based on traditional parameters such as the term frequency and the term distribution(idf) that are initially proposed in the field of IR to weight the documents and query terms in the VSM model. These techniques did not consider the semantic meaning of the text and produced low-quality summaries that hurt the recall of the developed IR system [8].

- Time penalty: the second reason that impedes the use of the ATS in the IR system is the time penalty of applying advanced statistical techniques that use the semantic analysis to summarize the documents such as the Latent Semantic Analysis (LSA) model. The LSA is a statistical model that can simulate the way people acquire knowledge and meaning through the correlation of facts from several sources [10]. The LSA is proposed in the literature to solve the VSM semantic problems [11], but the time consumption of the LSA is the challenge [12] [13].

An improved solution should be designed to be feasible for the IR indexing system (from the time and contents perspectives) and effective from the IR relevancy perspective (retrieve the desired relevant document).

To solve the two problems discussed in the previous paragraphs, we proposed to equip the IR system with the Multi-Layer Similarity (MLS) model. The MLS model is a text summarization model developed by Ababneh et al. in [14]. The model uses a multilayer approach of statistical analysis with a semantic investigation in complicated cases to generate a condensed version of the text document. The MLS model uses the LSA in the case that the verbatim similarity and the VSM similarity obtain low similarity results.

In summary, This research aims to improve the efficiency and performance of information retrieval systems. We will examine if the employment of the MLS model for text summarization can give the information retrieval systems the ability to obtain more relevant documents in a short period of time. The research achieves the following objectives:

- Improving the retrieval time through the reduction of the index size, which will be constructed from the summaries instead of the original documents.

- Proving that the use of the traditional statistical bag of word models ( such as the VSM and Jaccard coefficient) is not suitable for performing reasonable text summarization, especially to reduce the inverted index in an IR system.

- Using the IR evaluation measures to analyzing the relevancy of the Information Retrieval systems with and without MLS summarization.

The paper is organized into seven main sections. After discussing the importance of the research in section 1, section 2 reviews the important publications that addressed the use of automatic summaries to construct the inverted index in the IR systems. Section 3 explains how we employed the MLS model to construct a condensed inverted index and describes the IR system that was developed in this paper. Section 4 describes our experiments; the description includes the datasets and the relevancy measurements that were used to test the effectiveness of the employment of the MLS model in the IR field. Also, section 4 describes the collected results from our experiments. Section 5 presents the analysis of the results, and in section 6, we draw our evaluation. Section 7 presents the final achievements and the proposed future plan.

## II. RELATED WORK

The employment of automatic text summarization in the field of information retrieval is not a new topic. In [7], [8], and [9], the authors used the summaries that were generated from statistical models to build the inverted index. They used traditional statistical techniques to design their summarization models. These techniques depend on the term frequency and the term distribution(idf) over the text documents.

Ababneh et al. in [15] showed that modest effort had been made to measure the effect of the automatic text summarization in the performance of the IR systems. Brandow et al. in [7] obtained a high precision measuring rate when they used a domain-independent automatic summary (extractive summary) based on the traditional tf-idf sentence selection as an index source. They compared the results obtained from their extractive summary with another simple summary whose sentences are selected from the first few sentences in the original document (called Lead summary) and also with full-text indexing. The authors of [7] tested the relevancy against three condensation rate (CR) 160, 150, 250 words, and they obtained the highest precision at 150 and 250 CR values, but the recall decreased from 100% to 59%. Wasson in [16] used the first few sentences as the summary and built an inverted index from the summaries that were generated in this simple way. Wasson used news headlines and lists as a dataset, and found that the acceptability rate of the reader has been declined because the recall decreased to 23%. Note that the first attempts in this field either hurt the recall or obtained significant results at high value of the condensation rate.

---

[1] CR equals the summary length divided by the document length

The same conclusion can be driven from Sakai and Sparck-Jones in [8], they employed the same idea of using the summary for indexing, but they studied the impact of Generic summaries with or without Pseudo-relevance feedback. The results of Sakai and Sparck-Jones in [8] were in line with the results obtained in [7], in which a summary-only as a source of index obtained the most considerable precision at the highest condensation rate (see table 5 and 6 in [8]). Another essential thing that raised the precision in [8] is the inclusion of the document title in the index, and the title mainly conveys important terms or concepts.

Another employment of ATS in IR -especially in Geographical Information Retrieval subfield GIR – was done by Perea-Ortega et al. in [9]. They utilized two kind of summaries, General summary based on word frequency and noun phrases, and Geographic summary, which gave more attention to the sentences that contain Geographical entities. The authors in [9] gave a definite conclusion that the use of statistical single document summarizes as the source of indexing is not significant.

Recently, Ababneh et al. [14] developed a new semantic model for automatic text summarization. The model concentrates on the efficient use of the Latent Semantic Analysis (LSA). This model processes the text documents from two perspectives: statistically, by collected the essential statistical data from the text ( such as the term frequency, and the inverse term frequency), and semantically, by employing the LSA for concept-based text processing. The model was implemented and experimented as an ATS tool, but no research effort has been made to measure its effect on the relevancy in the IR field.

From the surveyed publications, we note the following three facts:

- in all the experiments that were performed in [7], [8], and [9], the precision was improved, and the recall was declined.
- All the surveyed research did not employ any semantic model, which goes beyond the traditional statistical models, in summarizing the text documents before initiating the indexing process.
- Ababneh et al. in [15] showed that modest effort had been made to measure the effect of the automatic text summarization in the performance of the IR systems. Furthermore, our related work section proves this point because we found only four publications that addressed the employment of ATS in the IR fields.

Based on these three facts, we chose to investigate this area of research that measures the effect of the semantic-based automatic text summarization on the relevancy and efficiency of the IR system. We chose the MLS model, which is a semantic text summarization model that showed significant accuracy results at a high condensation rate [14].

## III.    MLS-BASED RETRIEVAL METHOD

This section develops a new solution to satisfy the main aim specified in the introduction section. The aim is to improve information retrieval efficiency and performance. We seek to design a new method that remedies the efficiency problems that emerge from the large size inverted index in the information retrieval applications. Our method used efficient statistical and semantic models to reduce the size of the original inverted index to a short and informative inverted index. The method includes two phases; phase one includes the process of the documents by new, efficient, and accurate summarization model, and phase two designs an information retrieval system based on the VSM model [17].

The VSM is the most commonly used model in the IR field, and this is why we chose this model. We chose the model that already used and tested for a long time. Note that our target is to test the effect of the semantic text summarization in reducing the inverted index size and how this reduction affected the IR relevancy and space efficiency, not to improve the query/document matching models. Therefore, we used a well-known and tested IR model.

### A. General Architecture

Figure 1 gives a general view of the sequence of processing that was followed in our method. The input is a huge set of text documents. These documents comprise 40,006 documents that were taken from Essex, Kalimat, 242, and Blog Authorship datasets, as described in experiment section, and the output is a set of retrieved document $RD_1$, $RD_2$, $RD_3$,..., $Rd_i$ or $RD_1$, $RD_2$, $RD_3$,..., $Rd_j$ that are retrieved based on a VSM matching model. In the beginning, we want to mention that the dashed lines in Figure 1 follow the steps proposed in this research, and the solid lines follow the traditional steps in the IR system. The solid lines are inserted to the figure to mention that a comparison will be established between our method of retrieval and the traditional retrieval in the VSM model.

Figure 1 shows two inverted indexes, the Summary-based inverted index, and the Main Corpus (MC) inverted index. The Summary-based inverted index is the inverted index generated after summarising the original documents by the MLS model, and the MC inverted index is the inverted index of the original documents without summarization.

The architecture shown in Figure 1 presents the IR architecture with and without MLS text summarization. The VSM-Based IR system accepts two input data; the inverted index (either MC or summaries index) and the user query. The IR system makes the necessary matching, retrieving, and ranking. Note that in Figure 1, the value of i, which represents the number of documents retrieved in MC-based retrieval[2], and the value of j, which represents the number of documents retrieved in MLC-based retrieval[3], may or may not be identical. The two lists are collected for relevancy and efficiency comparison.

### B. Phase One: Summarizing the Text Documents Using the MLS Model

As appeared in [14], the intrinsic evaluation that has been performed of the MLS model over Kalimat data corpus showed significant results. The intrinsic evaluation considers the human summary as an ideal answer and compares the system generated summary against the human-generated summary to find the resemblance between them. Recall and precision are the primary measures of the intrinsic approach. In [14], the MLS model achieved 70% precision and 57% recall at a 42% condensation rate. Regarding the time efficiency improvement, the MLS model succeeded in reducing the number of runs of the LSA procedure by 52%. These results encourage us to use this model as the summarization tool because it produces significant accuracy in a reasonable time.

---

[2] The MC based retrieval is the retrieval model that use the main corpus as a source of the inverted index.
[3] The MLS based retrieval is the retrieval model that use the MLS summaries as a source of the inverted index.

The MLS model is a model for extracting generic summaries from the text documents by deleting the repetitive sentences in the document. The MLS model recognizes the repetitive sentences through a sophisticated statistical and semantic process. As shown in Figure 1, the MLS model integrated three text mining statistical techniques in a multi-layer summarization process. These techniques include the Jaccard Coefficient Model (JCM) in the first layer, the Vector Space Model (VSM) in the second layer, and the Latent Semantic analysis in the third layer [14].

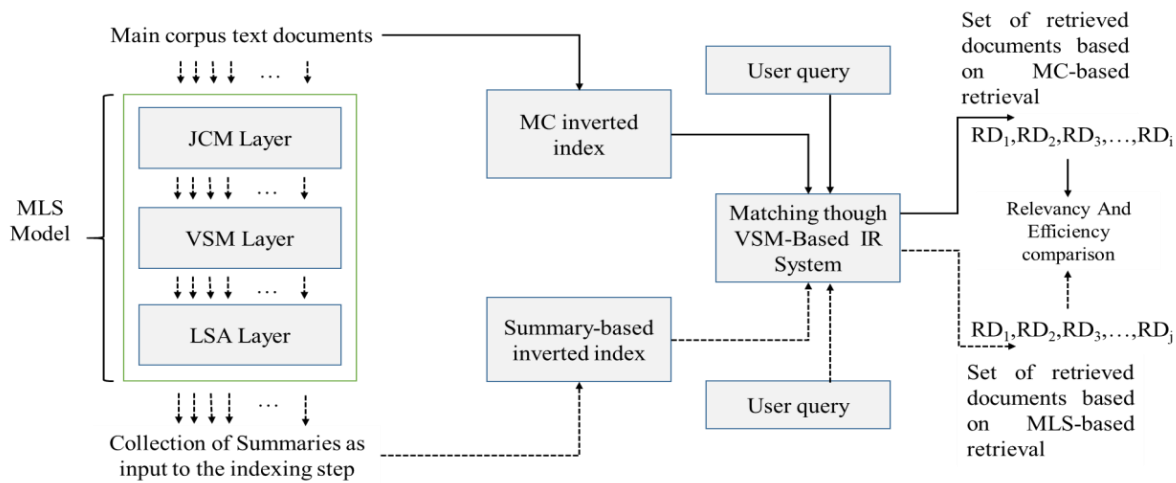

Figure 1. IR system architecture with MLS model

Figure 1 shows the sequence of execution in the MLS layer; it processes the text starting from the first layer, then the second layer, then the third layer. The purpose of this sequence of processing is to shift the time-consuming model (LSA model) to the upper layer and minimize the number of times we need it.

The JCM layer processes the verbatim similarity or the potion of the text that shares a significant number of terms' stems. The JCM layer in the MLS model uses the following similarity equation:

$$\mathbf{Jac}(\mathbf{S_i}, \mathbf{S_j}) = \frac{|\mathbf{Si} \cap \mathbf{Sj}|}{\min\limits_{\mathbf{S_i}, \mathbf{S_j} \neq \emptyset}(|\mathbf{S_i}|, |\mathbf{S_j}|)} \qquad (1)$$

*Where $S_i$ is the set of the stems found in sentence $i$, $S_j$ is the set of the stems found in sentence $j$*

The VSM layer process the portion of the text that has a significant value of tf4 and idf5. The VSM model in the MLS model represents each sentence in the document as a vector in the vector space. The vector contains the weights of the stems found in the sentence. Then, the VSM layer computes the cosine similarity between any two vectors:

$$\mathbf{cos}\left(\overline{\mathbf{s_i}}, \overline{\mathbf{s_j}}\right) = \frac{\sum_{n=1}^{t} \mathbf{w_{n,i}} \ \mathbf{w_{n,j}}}{\sqrt{\sum_{n=1}^{t} \mathbf{w_{n,i}}^2} \ \sqrt{\sum_{n=1}^{t} \mathbf{w_{n,j}}^2}} \qquad (2)$$

---

4 tf :term frequency in the text document
5 Idf: the number of sentences that contain the term in the document

*Where $\overline{s_i}$ is the vector of sentence $i$, $\overline{s_j}$ is the vector of sentence $j$, $w_{n,i}$ is the weight of term $n$ in sentence $i$, $w_{n,j}$ is the weight of term $n$ in sentence $j$, and $t$ is the number of terms in the document.*

The LSA layer processes the portion of the text the needs a semantic meaning understanding of the text being processed. In the hierarchal structure of the MLS model, the LSA model is used in the third layer, and the final equation of the MLS model implements this idea, as shown in equation 3:

$$\text{sim}(S_i, S_j) = \begin{cases} \dfrac{|S_i \cap S_j|}{\min\limits_{S_i, S_j \neq \emptyset}(|S_i|, |S_j|)}, & S_i, S_j \in X \\[2ex] \dfrac{\overline{S_i} \cdot \overline{S_j}}{|\overline{S_i}| \cdot |\overline{S_j}|} \; S_i, S_j \in X, \text{if } Jac(S_i, S_j) < 0.5 \\[2ex] \dfrac{\overline{S_i} \cdot \overline{S_j}}{\|\overline{S_i}\| \cdot \|\overline{S_j}\|} \; S_i, S_j \in S_q \cdot V_q^T, \text{if } Jac(S_i, S_j) \text{ and, } \cos(\overline{S_i}, \overline{S_j}) < 0.5 \end{cases} \quad (3)$$

*Where $S_i, S_j$ are any two sentences that belong to a text document D, and D could be a single document or multi-documents.*

Note that the LSA similarity, which represents the third branch in equation 3, is performed if the Jaccard and VSM similarities were insignificant.

The layer structure of the MLS boosts the time penalty of the model from two perspectives. Firstly, the MLS starts the processing of the text by using the JCM layer and then the VSM layer, those two layers are using efficient statistical techniques, as the authors in [14] proved. Secondly, the most time-consuming layer, which is the LSA layer, will not repeat the processing of the text portions that had already processed by the JCM and the VSM layers.

*C. Phase Two: IR System Based in the VSM Model*

In our method, we built an IR system based on the VSM model. The VSM was proposed in the field of IR by Salton [17]. The VSM is the most widely used model in information retrieval and natural language processing [18], [19], [20]. The retrieved set of documents from the VSM model is ranked according to the cosine similarity value, and the model allows the partial match.

The VSM is an algebraic model for matching documents and queries [2]. It has a robust mathematical foundation in which the documents and queries are depicted as vectors in multidimensional space. The components of each vector are a set of terms' weights that reflect the importance of these terms in the document.

$$\overline{d_j} = (w_{1,j}, w_{2,j}, w_{3,j}, \dots, w_{n,j})$$
$$\overline{Q} = (w_{1,q}, w_{2,q}, w_{3,q}, \dots, w_{n,q})$$

*where $\overrightarrow{d_j}$ is the vector of document j in the collection, $w_{1,j}$ is the weight of the term 1 in document j, $q_{1,j}$ is the weight of the term 1 in the query.*

An important issue that should be considering when we talk about VSM is the weighting scheme. No standard weighting scheme is found [2], but the best-known weighting scheme was proposed by Salton in [17], it is called the tf.idf weighting scheme, where $tf$ is the frequency of term $i$ and $idf$ is the number of documents that contain $i$,

$$w_{t,d} = (1 + \log f_{t,d}) \log \frac{N}{df\,t} \dots (4)$$

*Where $w_{t,d}$ is the weight of the term t in text d, $f_{t,d}$ is the frequency of the term t in text d, dft is the number of text segment contains t, N is the number of text segments in the corpus; text segment could be document or query.*

In the *tf.idf* weighting scheme, the terms that frequently appear in a particular document and distributed over a few numbers of documents take more weights than the terms that appear in every document. Thus, the stopwords and the general nouns and verbs that appear everywhere in the text and do not represent concepts or topics gain insignificant weights.

After computing the weights and preparing the documents' vectors, VSM calculates the similarity between each document and the user query by computing the cosine of the angle between the vectors that represent them [21].

$$\text{sim}\left(\overrightarrow{d_j}, \overrightarrow{Q}\right) = \cos\left(\overrightarrow{d_j}, \overrightarrow{Q}\right) = \frac{\overrightarrow{d_j} \cdot \overrightarrow{Q}}{|\overrightarrow{d_j}| \cdot |\overrightarrow{Q}|}$$

$$\text{sim}\left(\overrightarrow{d_j}, \overrightarrow{Q}\right) = \frac{\sum_{i=1}^{t} wd_{j_i} \cdot wQ_i}{\sqrt{\sum_{i=1}^{t} wd_{j_i}^2} \cdot \sqrt{\sum_{i=1}^{t} wQ_i^2}} \dots (5)$$

*Where $\overrightarrow{d_j}$ is the vector of document j, $\overrightarrow{Q}$ is the vector of query Q, $wd_{j_i}$ is the weight of the term i in $d_j$, $wQ_i$ is the weight of the term i in Q, t is the number of terms in the whole corpus*

### D. IR System Development

In our method, the VSM is returned to its origin since the development of the VSM was first proposed in the field of IR [17]. The theoretical background of the VSM model in IR is described in the previous section. In this paper, the IR model was developed based on the methodology described in phase 2 and implemented using VB 2013 programming language with excel sheets as interfaces. Table 1 presents the VB developed functions and the role of each one.

In our work, and the actual steps in designing the IR system came as follow:

**Step 1, Summarizing the documents:** As shown in Figure 1, the documents, which comprise our corpus, were summarized using the MLS model. Besides the MLS model, three other summarization models have been tested separately. These models include the LSA model, the VSM model, and the Jaccard model. We used models because they represent the different models that are integrated to form the MLS model. We used the software packages: MLSExtractor, LSAExtractor, VSMExtractor, and JacExtractor, which have been developed based on these four text summarization models [14].

The purpose of comparing the effect of the MLS model on the IR system with the effect of the other three models is to see if the MLS overtakes the existing models from the extrinsic[6] evaluation point of view. Thus, we collected the results of the relevancy measurements and the size of the inverted index of the IR systems before initiating the summarization and after applying the MLSExtractor, the LSAExtractor, the VSMExtractor, and the JacExtractor.

**Step 2, Inverted index creation:** in this step, we built several inverted indexes: MCECII which represents the inverted index of the main corpus (without summarization), MLSECII which represents the inverted index of the main corpus after summarizing its document using the MLS model, LSAECII which represents the inverted index of

---

[6] The extrinsic evaluation measures the effect of summarization tasks on the other field of natural languages processing and information retrieval fields.

the main corpus after summarizing its document using the LSA model, VSMECII which represents the inverted index of the main corpus after summarizing its document using the VSM model, JACECII which represents the inverted index of the main corpus after summarizing its document using the JCM model. The reason for generated several indexes is explained in the Experiments and Results section, but all of them are constructed using the same weighting scheme, and the same structure, as in the following form:

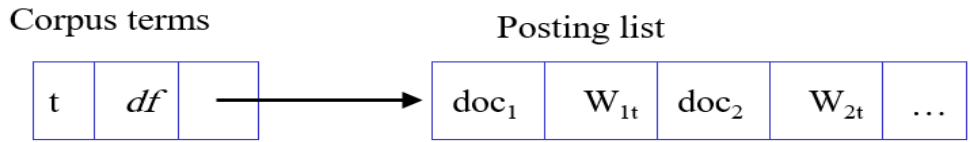

*Where t either represents the term found in the corpus of original documents (MC inverted index in Figure 1) or the term found in the corpus generated from the MLS model (Summary-based index in Figure 1), the df is the document frequency of the terms t or the number of documents that contains t, and the $W_{it}$ is the weight of term t in document i. The $W_{it}$ has been computed based on equation 4.*

**Step 3, Similarity computation:** after computing the weights and creating the inverted indexes, we used equation 5 to calculate the similarity between each document and the user query by computing the cosine of the angle between the vectors that represent them. Table 3 shows a sample of the similarity values.

**Step 4, Ranking the retrieved documents:** the system returns the documents that match the query based on the similarity computed in step 3. The retrieved set of documents is sorted in descending order based on the similarity values. Table 4 shows a sample of the retrieved list.

TABLE 1 VB FUNCTIONS CREATED IN THE IR SYSTEM

| VB function | Role |
|---|---|
| ExtractCorpusCreator | Constructs the summaries' corpus from the main corpus based on the sentences extracted from **MLSExtractor, LSAExtractor, VSMExtractor, and JacExtractor**) |
| InvertedIndexCreator | Creates the inverted index of the main corpus |
| ExtractInvertedIndexCreator | Creates the inverted index from the summaries corpus that is constructed by the **ExtractCorpusCreator** |
| DocumentLength | Finds the length of each document ( necessary for document length normalization) |
| TermWeight | Finds the weights of the terms. |
| DocLenNormalize | Normalizes the term weight based on the output of the **DocumentLength** |
| Ir | Finds the similarity between the query and all the documents in the main or summary corpus-based |
| RecallandPrecision | finds the recall and precision after the **Ir** running. |
| Inperppr | finds the Interpolated Average Precision after the **RecallandPrecision** running, and draws the precision-recall curve |
| MAP | finds the main average precision after the **RecallandPrecision** running. |

## IV. EXPERIMENTS AND RESULTS

A series of experiments with different conditions and constraints have been held. All the experiments were performed on Intel® Core™ i5-7200U CPU @ 2.5GHz processor with 8 GB RAM and Windows 10 OS.

### A. Experiment Datasets

The Arabic and English Languages have been chosen as the language of the case study. Four datasets have been chosen to perform our experiments:

- Essex Arabic Summaries Corpus: This corpus is published free on http://www.lancaster.ac.uk/staff/elhaj/corpora.htm. The corpus contains 153 Arabic articles and 765 human-generated extractive summaries. For each document, there are five manual summaries. The corpus contains documents with different subject areas, including art, music, science, technology, education, finance, health, politics, and religion. The corpus used recently by Al-Radaideh and Bataineh in [22].

- Kalimat data corpus[7] : Kalimat contains 20,291 Arabic article (3,537,677 Noun, 1,845,505 Verb, 115225 adjectives, and totally 6,286,217 terms). The corpus comprises greater than 6,000,000 terms. The data was taken from Omani newspapers. We tried to vary the topics and the domain of knowledge, so the selected data talking about health, science, history, art, religion, technology, environment, economic, and financial aspects.

- 242 data corpus, the corpus includes 242 Arabic text documents, 60 queries with their manual relevancy assessments. The corpus used by many researchers who investigated the Arabic IR field [23].

- The Blog Authorship corpus, it is an English language corpus that collects the posts of 19,320 bloggers and contains 681,288 text document composing 140 million words [24].

### B. IR Relevancy Measures Used in our Experiment

The evaluation of the information retrieval system is a complicated process. Measuring the effectiveness of the IR application means evaluating the relevancy, which is seen from the user's point of view. However, the researchers in these fields have developed a group of relevancy measures which become standard relevancy measurements for all IR system. These measures include Precision, Recall, F-score, and others [25]. Table 2 summarizes the IR relevancy measurements used in our experiments. The unranked retrieval measurements give a general indication about the contents of the retrieved set, whereas, the ranked retrieval measurements gives a strong indication about the quality of the answer set and magnifies the retrievals that rank the relevant documents on the top of the answer list [26].

### C. Experiment settings

Before proceeding, the following concepts have the following meanings in the explanation of the IR experiments.

- MC-based retrieval: the retrieval process that uses the main corpus to construct the inverted index (without summarization).

- MLS-based retrieval: the retrieval process that uses the summaries generated from the MLSextrcator to construct the inverted index.

---

[7] The corpus is available free on http://www.lancaster.ac.uk/staff/elhaj/corpora.htm

- LSA-based retrieval: the retrieval process that uses the summaries generated from the LSAextrcator to construct the inverted index.

- VSM-based retrieval: the retrieval process that uses the summaries generated from the VSMextrcator to construct the inverted index.

- JAC-based retrieval: the retrieval process that uses the summaries generated from the JACextrcator to construct the inverted index.

- The MLS-based retrieval, LSA-based retrieval, VSM-based retrieval, and JAC-based retrieval are called summary-based retrieval.

TABLE 2  IR EVALUATION MEASURES

| Precision (P) | *The number of relevant documents retrieved divided by the total number of retrieved documents.* | *Unranked Retrieval Evaluation* |
|---|---|---|
| Recall (R) | *The number of relevant documents retrieved divided by the total number of relevant documents.* | *Unranked Retrieval Evaluation* |
| F-Measure ( F ) | *The harmonic mean of R and P*<br>$F = \dfrac{2\ P\ R}{P\ +\ R}$ | *Unranked Retrieval Evaluation* |
| R-th Precision | *The precision value at a specific position in the retrieved ranked list. ( the number of the relevant document should be known in advance).* | *Ranked Retrieval Evaluation* |
| Average Precision ( AP ) | *Computed for each query, in which we compute the precision when each relevant document is retrieved, then we compute the average of all precision values obtained.* | *Ranked Retrieval Evaluation* |
| Mean Average Precision  (MAP) | *Computed for all queries, equals the average of  AP* | *Ranked Retrieval Evaluation* |
| Interpolated Average Precision | *Traces the maximum precision at 11 recall levels, Ri = {0.0, 0.1, 0.2 , 0.3, 0.4, 0.5, 0.6, 0.7, 0.8, 0.9, 1.0}, R0=0.0, R1=0.1... and R10=1.0*<br>$$P(R_i) = \max_{i \le r \le i+1} P(R)$$<br>*This measure answers the question; what is the maximum precision value achieved when the recall values ranged between x and y?* | *Ranked Retrieval Evaluation* |

The IR system experiments include the following experiments:

1) Experiment 1:

- Purpose: compare the relevancy results obtained from the MC-based retrieval with the relevancy results obtained from the summary-based retrieval when the relevant documents are selected manually for each query.

- The number of inverted indexes created is 5, one from the main corpus, one from the MLSExtractor summaries, one from the LSAExtractor summaries, one from VSMExtractor summaries, and one from the JacExtractor summaries.

- The number of processed queries is 60. The length of the queries distributed between 2 words such as „تمييز الاشكال بواسطة الحاسب الالي Information Technology تقنية المعلومات“ to five words such as „ shapes recognition by the computer.“

- Manual relevancy assessment performed by the corpus creators, for example for the query “ تقنية المعلومات” the following documents are recognized as relevant: 10, 96, 145, 175, 239

- Arabic language datasets

2) Experiment 2:

- Purpose: measure the convergence in the relevancy results for the summary-based retrieval to the relevancy results obtained from the MC-based retrieval.

- The number of inverted indexes created is 5, one from the main corpus, one from the MLSExtractor summaries, one from the LSAExtractor summaries, one from VSMExtractor summaries, and one from the JacExtractor summaries.

- The number of processed queries is 100.

- Automatic relevancy assessment, the retrieval list of the MC-based retrieval as the relevant list.

- Arabic language datasets

3) Experiment 3:

- Purpose: Measuring the effect of each summarization model on the relevancy of the IR system when the corpus is not semantically rich (the text's writers do not diversify their vocabularies) . to achieve this purpose we used the Blog Authorship corpus which represents young people posts, and usually those people in their posts do not diversify their vocabularies.

- The number of inverted indexes created is 5, one from the main corpus, one from the MLSExtractor summaries, one from the LSAExtractor summaries, one from VSMExtractor summaries, and one from the JacExtractor summaries.

- The number of processed queries is 60.

- Automatic relevancy assessment, the retrieval list of the main corpus inverted index as the relevant list.

- English language datasets.

The number of inverted indexes created in each experiment is five as follows:

- **MCII**: Main corpus inverted index

- **MLSECII**: Inverted index of the MLS summaries corpus, the inverted index created based on the summaries that were generated from the MLSExtractor.

- **LSAECII**: Inverted index of the LSA summaries corpus, the inverted index created based on the summaries that were generated from the LSAExtractor

- **VSMECII**: Inverted index of the VSM summaries corpus, the inverted index created based on the summaries that were generated from the VSMExtractor

- **JACECII**: Inverted index of the Jaccard summaries corpus, the inverted index created based on the summaries that were generated from the JacExtractor

The output from each experiment includes:

1) The similarity values between the queries and the documents found in the inverted index as follow:

- Sim(MC, Q): the similarity between the queries and the documents appeared in **MCII**.

- Sim(MLSE, Q): the similarity between the queries and the documents appeared in **MLSECII.**

- Sim(LSAE, Q): the similarity between the queries and the documents appeared in **LSAECII.**

- Sim(VSME, Q): the similarity between the queries and the documents appeared in **VSMECII.**
- Sim(JACE, Q): the similarity between the queries and the documents appeared in **JACECII.**

2) The precision at each point a relevant document is retrieved (for each query and each inverted index), for example, in Experiment 3, we produced the followings:

- **RP(MLS,60, MC)** → MLSECII inverted index, 60 queries, main corpus retrieved set as a relevant list.

  *This represents the precision at each retrieve of relevant document for 60 queries, the MLSECII is the inverted index, and the relevancy assessment is based on the retrieved set of running the IR system over the main corpus inverted index (in simple words comparing the results of running the IR system over the main corpus inverted index and over the MLSECII inverted index, and the comparison includes 60 queries).*

- **RP(LSA,60, MC)** → LSAECII inverted index, 60 queries, main corpus retrieved set as a relevant list.

  *This represents the precision at each retrieve of a relevant document for 60 queries, the LSAECII is the inverted index, and the relevancy assessment is based on the retrieved set of running the IR system over the main corpus inverted index.*

- **RP(VSM,60, MC)** → VSMECII inverted index, 60 queries, main corpus retrieved set as a relevant list.

  *This represents the precision at each retrieve of a relevant document for 60 queries, the VSMECII is the inverted index, and the relevancy assessment is based on the retrieved set of running the IR system over the main corpus inverted index.*

- **RP(JAC,60, MC)** → JACECII inverted index, 60 queries, main corpus retrieved set as a relevant list.

  *This represents the precision at each retrieve of a relevant document for 60 queries, the JACECII is the inverted index, and the relevancy assessment is based on the retrieved set of running the IR system over the main corpus inverted index.*

  **(In Experiment 2 the number 60 is replaced by 100 because the number of processed queries is 100, and in Experiment 1 the MC is replaced by Manual because the relevancy assessment is performed manually and already found in the dataset)**

3) The final recall value of the IR system for each inverted index.

4) The final MAP value of the IR system for each inverted index.

5) The size of each inverted index.

6) The ratio of the size of the inverted index to the main corpus inverted index.

7) The Interpolated Average Precision at 11 recall points of the IR system for each inverted index.

8) The Precision-Recall curve of the IR system for each inverted index.

*D. Experiments Results*

As described previously, three experiments were conducted to measure the effect of different models of text summarization (Jaccard, VSM, LSA, and MLS) on the size of the inverted index, and on the relevancy of the IR system. To unify the way we judge the effect of these models, we collected the following results from all the experiments:

**The similarity values between the queries and the documents (Experiment 1 to Experiment 3):** the similarity values between each query and the documents, which were represented in five inverted indexes, have been collected. The similarity values are used to retrieve the set of relevant documents. Table 3 represents the similarity values obtained in Experiment 1 for the query "هندسة الحاسوب Computer Engineering" based on the five inverted indexes created in Experiment 1:

TABLE 3: THE SIMILARITY VALUES OBTAINED IN EXPERIMENT 1 FOR THE QUERY "هندسة الحاسوب COMPUTER ENGINEERING"-EXPERIMENT 1

| MC-based Retrieval | | MLS-based Retrieval | | LSA-Based Retrieval | | VSM-Based Retrieval | | JAC-Based Retrieval | |
|---|---|---|---|---|---|---|---|---|---|
| Doc id | Sim(doc,q) | Doc id | Sim(doc,q) | Doc id | Sim(doc,q) | Doc id | Sim(doc,q) | Doc id | Sim(doc,q) |
| 14 | 0.1980207 | 34 | 0.210828191 | 14 | 0.229657346 | 14 | 0.211339244 | 14 | 0.1980207 |
| 47 | 0.186255837 | 47 | 0.206865898 | 178 | 0.218801136 | 47 | 0.199591737 | 47 | 0.186255837 |
| 156 | 0.186065978 | 14 | 0.193870283 | 34 | 0.208062795 | 203 | 0.187697058 | 156 | 0.186065978 |
| 203 | 0.185606874 | 305 | 0.190996025 | 47 | 0.2051011 | 20 | 0.170028502 | 203 | 0.185606874 |
| 178 | 0.176122605 | 20 | 0.180476029 | 305 | 0.187611732 | 164 | 0.168159261 | 178 | 0.176122605 |
| 164 | 0.168637281 | 49 | 0.174686353 | 20 | 0.179086041 | 34 | 0.164408222 | 164 | 0.168637281 |
| 20 | 0.168309015 | 164 | 0.168763246 | 49 | 0.16907098 | 305 | 0.163394407 | 20 | 0.168309015 |
| 49 | 0.165555915 | 366 | 0.168312197 | 164 | 0.167083439 | 49 | 0.158704569 | 49 | 0.165555915 |
| 305 | 0.162653219 | 203 | 0.088592905 | 393 | 0.166200259 | 366 | 0.156211881 | 305 | 0.162653219 |
| 34 | 0.159897703 | 21 | 0.053575267 | 366 | 0.166101684 | 21 | 0.049498976 | 34 | 0.159897703 |
| 21 | 0.047744886 | 178 | 0.052969566 | 203 | 0.086123997 | 174 | 0.049093251 | 21 | 0.047744886 |
| 174 | 0.047528632 | 175 | 0.052670559 | 21 | 0.056814777 | 668 | 0.042377292 | 174 | 0.047528632 |
| 175 | 0.042801034 | 174 | 0.052581104 | 174 | 0.05220073 | 393 | 0.038545231 | 175 | 0.042801034 |
| 668 | 0.039167393 | 668 | 0.05201775 | 668 | 0.047259432 | 178 | 0.03347113 | 668 | 0.039167393 |
| 234 | 0.035457636 | 234 | 0.041072802 | 175 | 0.046668534 | 56 | 0.032582967 | 234 | 0.035457636 |
| 696 | 0.026202767 | 56 | 0.036293896 | 56 | 0.032032625 | 175 | 0.031920477 | 696 | 0.026202767 |

**The sets of retrieved documents for each query that were retrieved based on the similarity values between the query and documents (Experiment 1 to Experiment 3):** the retrieved sets of documents are sorted based on the value of similarity. The sets are then used to measure the relevancy measurements (Precision, recall, …) based on manual relevancy assessment found in the corpora described in the previous section or based on the relevant documents retrieved from MC-based retrieval. Table 4 represents the retrieved sets of documents that match the query " الاكتئاب و القلق Depression and anxiety" in the five inverted indexes that were created in Experiment 2.

**The sets of relevant documents for each query (Experiment 1 to Experiment 3):** the relevant documents in Experiment 1 are prepared manually in the employed datasets for example for the query "هندسة الحاسوب Computer Engineering" the set of relevant documents are 14, 24, 53, 71, 72, 75, 77, 93, 103, 178, 179, 180, 181, 182, 183, 184, 185, 186, 203, 216, 218, 219, 230. In Experiment 2 and Experiment 3, we took the retrieved documents based on the MC-based retrieval as the relevant documents. The purpose was to compare the relevancy that was achieved from the summary-based retrieval with the relevancy achieved from the MC-based retrieval. For example, the retrieved sets of documents that match the query "الاكتئاب و القلق Depression and anxiety" in the MC-based retrieval tested in Experiment 2 came as follow: 317, 203, 294, 328, 356, 377, 372, 702, 382, 292, 335, 693, 359, 373. And, the retrieved sets of documents that match the query "victims and criminals" in the MC-based retrieval tested in Experiment 3 (the English language corpus) came as follow: 35, 16, 20, 92, 214, 74, 263, 73, 219, 75, 159, 213.

TABLE 4: THE RETRIEVED SETS OF DOCUMENTS OF THE QUERY "الاكتئاب و القلق DEPRESSION AND ANXIETY" - EXPERIMENT 2

| MC-based retrieval | MLS-based retrieval | LSA-based retrieval | VSM-based retrieval | Jac-based retrieval |
|---|---|---|---|---|
| 317 | 294 | 317 | 203 | 317 |
| 203 | 335 | 294 | 377 | 203 |
| 294 | 373 | 702 | 294 | 294 |
| 328 | 377 | 359 | 335 | 292 |
| 356 | 702 | 372 | 292 | 372 |
| 377 | 382 | 382 | 372 | 328 |
| 372 | 372 | 377 | 382 | 377 |
| 702 | 359 | 693 | 702 | 382 |
| 382 | 693 | 317 | 373 | 702 |
| 292 | | | 356 | 335 |
| 335 | | | 328 | 373 |
| 693 | | | 693 | 359 |
| 359 | | | 359 | 693 |
| 373 | | | | 356 |

**The precision at each retrieve of a relevant document((Experiment 1 to Experiment 3).** This measure is vital to generate the recall-precision cure. Table 5 was taken from the results of Experiment 1, and it represents the precision at each retrieve of a relevant document for the query id3 with MLS-based retrieval.

TABLE 5: THE PRECISION AT EACH RETRIEVE OF RELEVANT DOCUMENT FOR THE QUERY id3-EXPERIMENT 1

| Query id | Doc Ret | R | P |
|---|---|---|---|
| 3 | 203 | 0.0625 | 0.5 |
| 3 | 178 | 0.125 | 0.5 |
| 3 | 181 | 0.1875 | 0.107143 |
| 3 | 182 | 0.25 | 0.137931 |
| 3 | 186 | 0.3125 | 0.166667 |
| 3 | 179 | 0.375 | 0.181818 |
| 3 | 183 | 0.4375 | 0.179487 |
| 3 | 180 | 0.5 | 0.163265 |
| 3 | 207 | 0.5625 | 0.157895 |
| 3 | 184 | 0.625 | 0.138889 |
| 3 | 209 | 0.6875 | 0.127907 |
| 3 | 215 | 0.75 | 0.129032 |

**Interpolated Average Precision (Experiment 1 to Experiment 3):** this measure traces the maximum precision at 11 recall levels, Ri = {0.0, 0.1, 0.2 , 0.3, 0.4, 0.5, 0.6, 0.7, 0.8, 0.9, 1.0}. After computing the Interpolated Average Precision for all the queries, the average is computed for each interval. From Experiment 1, Table 6 represents the Interpolated Average Precision values for the query "شبكات الحاسب الالي Computer Networks" with the MLS-based retrieval:

TABLE 6: THE INTERPOLATED AVERAGE PRECISION VALUES FOR THE QUERY "شبكات الحاسب الالي COMPUTER NETWORKS" WITH THE MLS-BASED RETRIEVAL-EXPERIMENT 1

| Recall Level | P |
|---|---|
| r0 | 1 |
| r1 | 0.9 |
| r2 | 0.645833 |
| r3 | 0.651515 |
| r4 | 0.527864 |
| r5 | 0.351073 |
| r6 | 0.220784 |
| r7 | 0.179173 |
| r8 | 0.128708 |
| r9 | 0.103261 |
| r10 | 0 |

The 0.9 percent appeared with r1 means that the maximum precision obtained when the recall value was greater than or equal to 0.1 and less than 0.2 is 90%. Table 7 shows the average of the Interpolated Average Precision for all the queries (60 queries) in Experiment 1 and Experiment 2:

TABLE 7: THE AVERAGE OF THE INTERPOLATED AVERAGE PRECISION FOR ALL THE QUERIES (60 QUERIES) IN EXPERIMENT 1 AND EXPERIMENT 2

| Recall Level | AP Experiment 1 | AP Experiment 2 |
|---|---|---|
| r0 | 0.5086407 | 0.95049505 |
| r1 | 0.38638365 | 0.98019802 |
| r2 | 0.29987006 | 0.95049505 |
| r3 | 0.2320163 | 0.98019802 |
| r4 | 0.18198353 | 0.940594059 |
| r5 | 0.15292554 | 0.900990099 |
| r6 | 0.12920146 | 0.643564356 |
| r7 | 0.10158577 | 0.485148515 |
| r8 | 0.05960479 | 0.158415842 |
| r9 | 0.01138984 | 0 |
| r10 | 0 | 0 |

Note that the Interpolated Average Precision for Experiment 2 is higher than the Interpolated Average Precision for Experiment 1 because in Experiment 2 the relevancy is judge against the set of documents retrieved based on the main corpus inverted index, and these high values mean how close the summary-based retrieval to the main corpus retrieval. In the example above and at r5 ( recall between 0.4 and less than 0.5), the retrieved set contains 90% relevant documents of the documents retrieved when the IR system used the main corpus as the source of indexing.

**The Mean Average Precision (MAP):** as mentioned in Table 2, the average precision is computed when each relevant document is retrieved. The MAP equals the average of the average precision for all the queries. The importance of this measure is related to the quality of the retrieved set, and the retrieved set depends on the IR system used. If the retrieved set contains a sufficient number of relevant documents and the relevant documents appeared at the top of the retrieved list, then the value of the MAP will be high. Note that we unify the IR system, and we change only the source of the index, so we are not concerns about the values of the MAP; we concern about the convergence between the MAP of the MC-based retrieval and the MAP of the summary-based retrieval. This means that in case the IR system used the main corpus inverted index or the MLS summaries inverted index, the MAP value should be convergent (high or low this is not important). Table 8 presents the MAP that was obtained in Experiment 1.

**The Recall:** the recall measures the percent of relevant retrieved documents to the total number of relevant documents. The recall is significant because we want to measure the number of the relevant documents retrieved in the MC-based retrieval and the summary-based retrieval. Table 9 shows the obtained Recall values in Experiment 1:

TABLE 8: THE OBTAINED MAP IN EXPERIMENT 1

|     | MC-based Retrieval | MLS-based Retrieval | LSA-Based Retrieval | VSM-Based Retrieval | JAC-Based Retrieval |
| --- | --- | --- | --- | --- | --- |
| MAP | 0.399261778 | 0.368824711 | 0.371919606 | 0.385096685 | 0.393788301 |

TABLE 9: THE OBTAINED RECALL IN EXPERIMENT 1

|     | MC-based Retrieval | MLS-based Retrieval | LSA-Based Retrieval | VSM-Based Retrieval | JAC-Based Retrieval |
| --- | --- | --- | --- | --- | --- |
| Recall | 0.780724898 | 0.650635746 | 0.669158738 | 0.735720636 | 0.738503713 |

**The ratio of the summaries inverted index size to the main corpus inverted index size:** the sizes of the inverted indexes generated from the summaries that are produced from the four summarization models are measured. We measured the number of terms composing the inverted index. We did not measure the size in bytes because the accurate size is affected by the type of compression techniques, and it is out of our interest. We want to see the reduction in the size of the inverted index and how this affected the relevancy. Table 10 represents the ratio of size reduction in Experiment 1:

TABLE 10: THE RATIO OF THE SIZE REDUCTION – EXPERIMENT 1

|     | MC-based Retrieval | MLS-based Retrieval | LSA-Based Retrieval | VSM-Based Retrieval | JAC-Based Retrieval |
| --- | --- | --- | --- | --- | --- |
| Ratio to the main corpus inverted index | 100% | 42% | 54% | 68% | 79% |

All these results are collected from the three experiments for further analysis and evaluation. However, it is important to note that the sizes of the inverted indexes in Experiments 1, 2 are identical because we used the same inverted indexes but with different constraints (such as in Experiment 2, we increased the number of queries to 100). Only Experiment 3 has different sizes of the inverted indexes because they represent a new dataset (English Language dataset).

## V. ANALYSIS AND DISCUSSION

Experiments 1, 2, and 3 test the employment of different ATS techniques in reducing the inverted index and how this reduction affected the relevancy assessment. The kinds of collected results that are described in the previous section include the relevancy results (recall, precision, Interpolated Average Precision, MAP) and the inverted index size. The size of the inverted index is used to measure the enhancement achieved on the IR system performance. During the evaluation, we linked the recall-precision curve with the size of the inverted index and the final recall obtained at the end of each experiment.

In this section, the abbreviations appear in the figures have the following meanings: MC-curve refers to the recall-precision curve of the IR system that uses the main corpus to build the inverted index ( without summarization), MLS-curve refers to the recall-precision curve of the IR system that uses the summaries that were generated from the MLSExtractor, LSA-curve refers to the recall-precision curve of the IR system that uses the summaries that were generated from the LSAExtractor, VSM-curve refers to the recall-precision curve of the IR system that uses the summaries that were generated from the VSMExtractor, and JAC-curve refers to the recall-precision curve of the IR system that uses the summaries that were generated from the JacExtractor.

### A. Experiment-one Final Findings

Figure 2 shows the recall-precision curves that were obtained in Experiment 1. The curves trace the precision behavior at 11 recall points. The red curve represents the MC-based retrieval, and the other curves represent the summary-based retrievals. In Figure 2, the red curve represents the optimal relevancy results that were generated from the IR system. Note the slight differences between the red curve and the other curves that means that all the summary-based retrievals results succeeded in retrieving a considerable number of relevant documents. The LSA-curve and MLS-curve show a small drop compared with the JAC-curve and VSM-curve, especially after r5.

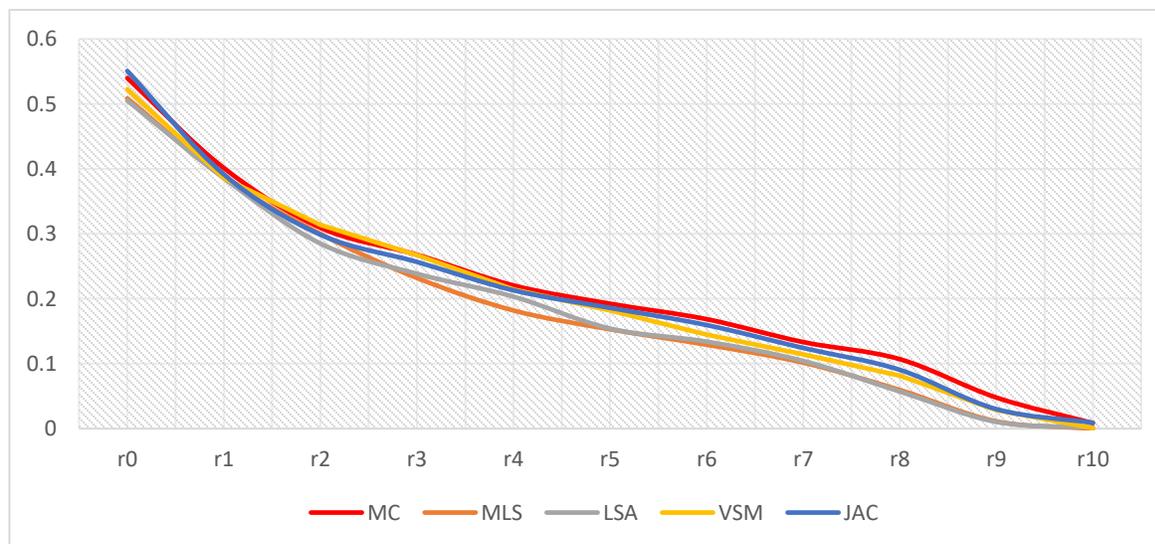

Figure 2. the Recall-Precision Curves in Experiment 1

The trend that appears in Figure 2 should be supplemented with the other results obtained in Experiment 1; the inverted index size and the final recall and MAP. Figure 3 presents the final results of the recall, MAP, and the ratio of the summaries inverted index size relative to the main corpus inverted index. Note that the MLS-based retrieval obtained convergent MAP results with the other summary-based retrievals, and note that the recall value of the MLS-based retrieval was less than the MC-based retrieval by 12%. The recall value of the MLS-based retrieval is 65%, which represents a reasonable result because we obtained it at 58% reduction in the inverted index. The JAC-based retrieval and VSM-based retrieval relevancy results were very close to the MC-based retrieval results but with inconsiderable reductions in the inverted index size (21%, 32%, respectively).

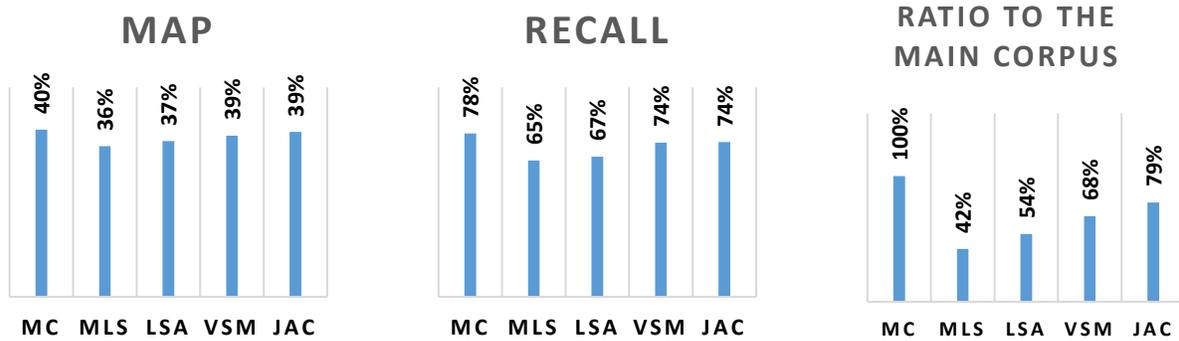

Figure 3. MAP, Recall, and the Ratio to the Main Corpus Size (in Experiment 1).

B. *Experiment-two Final Findings*

Comparing with Experiment 1, two changes were made In Experiment 2: 1) the number of queries was increased to 100, and the purpose was to test the behavior of the IR system with a larger number of queries, 2) the answer set retrieved from the MC-based retrieval was taken as the relevant set of documents. The inverted indexes in Experiment 1 and Experiment 2 are the same. Figure 4 shows the recall-precision curves obtained from Experiment 2.

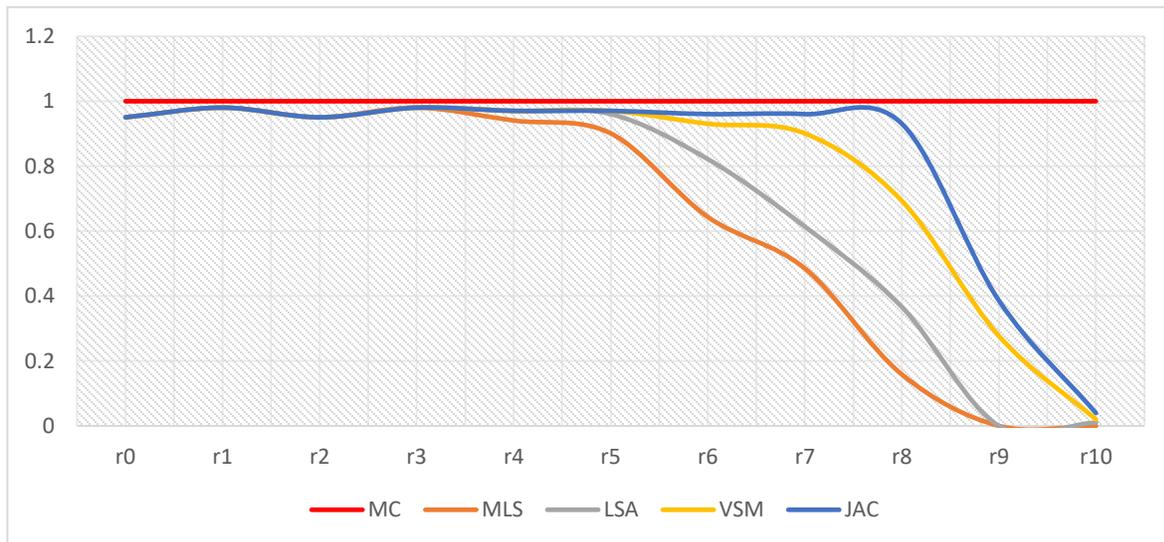

Figure 4. the Recall-Precision Curves in Experiment 2

Note that the MC red line always one because we considered it as the gold answer. From r0 to r5, all the summaries based retrievals' relevancy results were very close to the MC-based retrieval relevancy results, after r6 the summaries based retrievals starts to drop, and the most significant drop happened to the MLS-based retrieval.

The trend that appears in Figure 4 should be supplemented with the other results obtained in Experiment 2; the inverted index size and the final recall and MAP. Figure 5 presents the final results of the recall, MAP, and the ratio of the summaries inverted index size relative to the main corpus inverted index. Note that the MAP was very high for the four summary-based retrievals, and this reflects that the summary-based retrievals obtained reasonable precision. The recall behavior is the same as the one that appeared in Figure 3; the MLS-based retrieval obtained 63% recall value at 58% inverted index reduction, the recall of the VSM-based retrieval and JAC- based retrieval was above 80%, but the amount of reduction was inconsiderable. The relevancy assessment of Experiment 1 and Experiment 2 are roughly the same, the summaries based retrieval curves are very close to the curve obtained in the MC retrieval, and the MAP and recall have the same behavior.

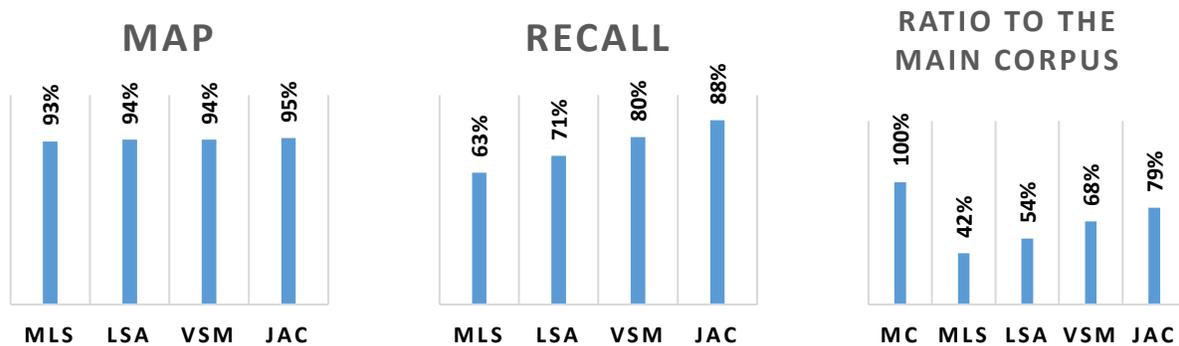

Figure 5. MAP, Recall, and the Ratio to the Main Corpus Size (in Experiment 2).

### C. Experiment-three Final Findings

In Experiment 3, an English language corpus was used instead of the Arabic Language corpus. The purpose of performing the Experiment 3 is to measure the effect of each model on the relevancy of the IR system when the corpus is not semantically rich (the text's writers do not diversify their vocabularies). Figure 6 shows the recall-precision curves obtained in Experiment 3. From r0 to r4, all the summaries based retrievals' relevancy results were very close to the MC-based retrieval relevancy results, after r5 the summaries based retrievals starts to drop, and the most noticeable drop happened to the MLS-based retrieval and the VSM-based retrieval. In this experiment, the VSM-based retrieval obtained fewer precision values than the LSA-based retrieval because the used corpus contains the posts of young people bloggers who typically do not diversify their vocabularies during the posting. This feature in the English corpus magnified the role of the second layer in the MLS summarization (VSM layer) and caused the VSM model to delete a large portion of the text based on simple statistical calculations (the role of semantic analysis is weak in this case). Also, this feature affected the precision values of the MLS-based retrieval because the MLS model uses the VSM model in the middle layer. The drop in the VSM-curve caused the drop of the MLS-curve, and it was not a drop that is based on the semantic analysis of the text.

In Figure 7, we see that the recall of the VSM-based retrieval and MLS-based retrieval is 54% and 55%, respectively. Also, the MLS-based retrieval and VSM-based retrieval showed 39% and 36% reduction in the inverted index size, which is not the desirable ratio. The obtained results in Experiment 3 proves one of our claims that the simple statistical analysis of the text analysis based on the term frequency and term distribution hurts the summarization results, and this affected the relevancy of the IR system that uses an inverted index that is reduced by the VSM model such as the IR systems appeared in [27], [7], [8].

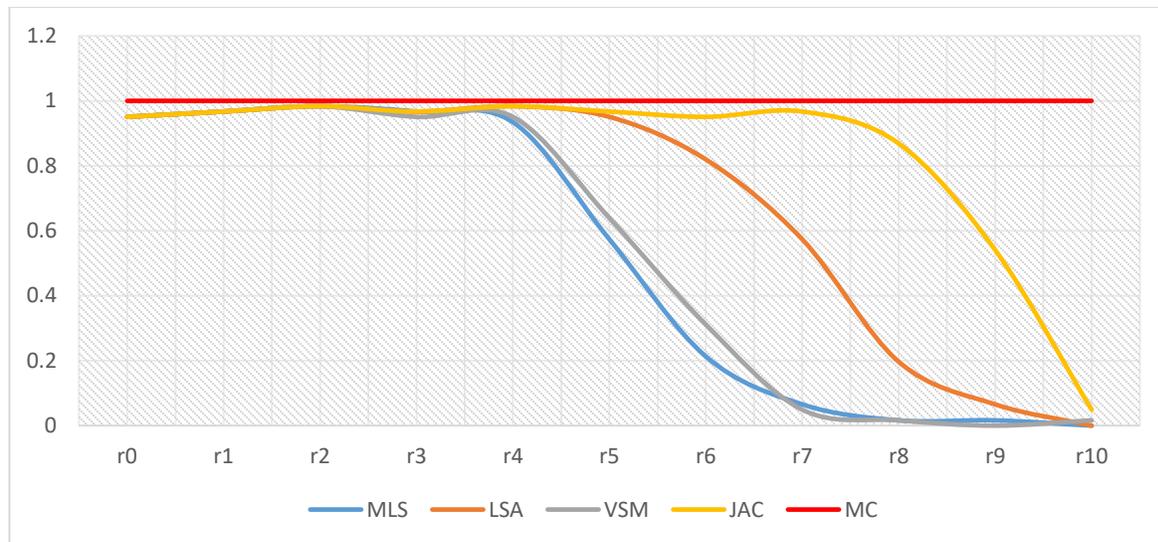

Figure 6. the Recall-Precision Curves in Experiment 3

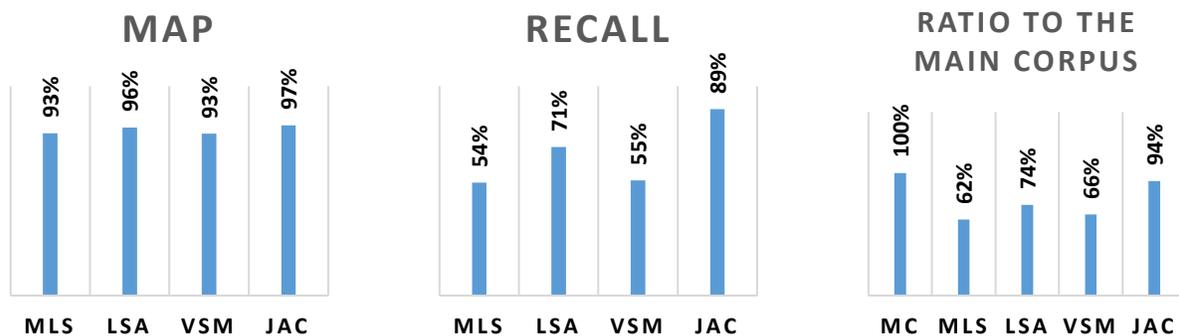

Figure 7. MAP, Recall, and the Ratio to the Main Corpus Size (in Experiment 3).

## VI. EVALUATION

The relevancy and efficiency analysis of the MLS-based retrieval results showed significant merits. We can note that the overall performance is significant as follow:

**Condense version of the main corpus inverted index:** The inverted index of the MLS based retrieval was less than the inverted index of the main corpus by 58% in Experiment 1 and 2, and by 38% in Experiment 3. Note that the sizes of the inverted indexes in the VSM and JAC-based retrievals were large and close to the inverted index size in the MC-based retrieval.

**Convergent relevancy results: a**s shown in the figures of the previous section, the MAP and the recall-precision curve of the MLS-based retrieval were very close to their counterparts in the MC-based retrieval. For example, in Experiment 1, the MAP of the MC-based retrieval was 40%, and the MAP of the MLS-based retrieval was 36%. In Experiment 2, the MAP of the MLS-based retrieval obtained 93% of the MAP of the MC-based, as shown in Figure 5. The recall-precision curves of Figures 2, 4, 6 are convergent until r5.

**Convergent relevancy results between MLS-based retrieval and LSA-based retrieval:** As shown in Figures 2, 3, 4, 5, the recall, MAP, and the recall-precision curve of the MLS-based retrieval were very close to their counterparts in the LSA-based retrieval. These relevancy results are obtained at a ratio to the main corpus equals 42% in MLS-based retrieval and 54% in LSA-based retrieval.

The recall value of the MLS-based retrieval was the lowest in all the IR experiments (Experiment 1-Experiment 3). However, it represents significant recall: as shown in Figures 3 and 5, the recall values of the MLS-based retrieval did not exceed r6 (it was less than 70%), whereas the LSA-based retrieval reached r7 in Experiment 3 and JAC-based retrieval reached r8 in the experiments 2 and 3. However, in the previous work, which was discussed in the introduction section, the precision was significant but the recall was clearly insignificant. For example, in [7], the recall value declined by 41% (from 100% to 59%) and in [9] the significant recall results obtained at 60%, 80%, and 90% condensation rate which implies that the reduction in the inverted index does not exceed 40%. In our work, we obtained high recall that reached 65% in Experiment 1 and the difference in the recall between the MC-based retrieval and the MLS-based retrieval does not exceed 13%, with a 58% reduction in the inverted index size.

However, the extrinsic evaluation of the MLS-based retrieval shows one problem that the intrinsic evaluation, which was performed in [14], did not discover. We found that **the role of each layer in the MLS text summarization is corpus dependent**. This means that the part of the text that should be processed in each layer depends on the diversity of the vocabularies that were used to build the corpus. This drawback affected the IR relevancy results. In Experiment 3, the vocabularies do not contain the required diversity because the dataset represents the young people's writings who usually do not diversify their vocabularies. This merit magnifies the role of the VSM layer because the value of the term frequency and inverse term frequency were, in most cases, considerable. So, the second layer (the VSM layer) had the greatest effect in the MLS summarization, and this caused to decrease in the role of the semantic layer (the LSA layer). Figures 6 and 7 showed the results of Experiment 3, the MLS-based retrieval and the VSM-based retrieval had convergent relevancy results and this contradicts the other experiments' results, which showed convergent results between the MLS-based retrieval and LSA-based retrieval.

## VII. CONCLUSION AND FUTURE WORK

### A. Achievements

The new method, MLS-based retrieval, has been developed and evaluated with three Arabic language datasets and one English dataset on the IR systems.

**The MLS-based retrieval method has a positive influence on the efficiency of the IR system without noticeable loss in the precision results.** The size of the MLS inverted index is 58% smaller than the size of the original documents inverted index and 12% smaller than the size of the LSA-based retrieval inverted index, which

implies less time to match the index and the query terms and less space to store the index in the main memory and the secondary storage devices.

**The precision relevancy measure in the three experiments that test the employment of the MLS in the IR system shows the convergence between the MLS-based retrieval and the MC-based retrieval.** The MAP of the MLS-based retrieval obtained 93% of the MAP obtained in the MC-based retrieval, and the recall-precision curves in the three experiments showed that the two curves that represent the MLS-based and MC-based retrievals were very close and the noticed difference appeared at high recall values (r6).

**The Recall is significant:** the results show a slight drop compared with the MC-based retrieval. However, we consider the MLS-based retrieval recall is significant because it represents 84% of the recall that was obtained in the MC-based retrieval.

Based on the recall, precision, and the size of the inverted index values of the MLS-based retrieval, we conclude that the inverted index, which has been built according to the summaries that were generated according to the MLS model, is **condensed and informative**.

### B. Future Work

The paper presented a solution that enhances the efficiency of the IR systems and, at the same time, maintains the relevancy of the retrieved list. In the future, we plan to make further efficiency improvement by enhancing the weighting scheme in the VSM model and replace it with a semantic-based and efficient weighting scheme. Besides considering the statistical appearance of the term, the new weighting scheme will link the term with its semantic context to determine the meaning of the term.